\documentclass[sigconf,natbib]{acmart}
\usepackage{multirow}
\usepackage{url}
\usepackage[normalem]{ulem}
\useunder{\uline}{\ul}{}
\AtBeginDocument{%
  \providecommand\BibTeX{{%
    \normalfont B\kern-0.5em{\scshape i\kern-0.25em b}\kern-0.8em\TeX}}}

\copyrightyear{2023}
\acmYear{2023}
\setcopyright{acmcopyright}\acmConference[SIGIR '23]{Proceedings of the 45th International ACM SIGIR Conference on Research and Development in Information Retrieval}{July 11--15, 2022}{Madrid, Spain}
\acmBooktitle{Proceedings of the 45th International ACM SIGIR Conference on Research and Development in Information Retrieval (SIGIR '23), July 11--15, 2022, Madrid, Spain}
\acmPrice{15.00}
\acmDOI{xx.xxxx/xxxxxx.xxxxxxxx}
\acmISBN{xxx-x-xxxx-xxxx-x/xx/xx}

\begin{document}

\title{Both Efficiency and Effectiveness! A Large Scale Pre-ranking Framework in Search System}

\author{Qihang Zhao}
\authornotemark[1]
\affiliation{University of Science and Technology of China, Anhui\country{China}}
\affiliation{Kuaishou, Beijing\country{China}}
\email{zhaoqihang@kuaishou.com}
\author{Rui-jie Zhu}
\authornotemark[1]
\affiliation{%
  University of California, Santa Cruz
  \country{California, USA}}
\email{ridger@live.cn}
\author{Liu Yang}
\authornotemark[1]
\affiliation{Kuaishou, Beijing\country{China}}
\email{liuyang22@kuaishou.com}
\author{He Yongming}
\affiliation{Kuaishou, Beijing\country{China}}
\email{heyongming@kuaishou.com}
\author{Bo Zhou}
\affiliation{Kuaishou, Beijing\country{China}}
\email{yefeng@kuaishou.com}
\author{Luo Cheng}
\affiliation{Kuaishou, Beijing\country{China}}
\email{luocheng06@kuaishou.com}
\authornote{corresponding author}




\renewcommand{\shortauthors}{Zhao, et al.}

\begin{abstract}
  In the realm of search systems, multi-stage cascade architecture is a prevalent method, typically consisting of sequential modules such as matching, pre-ranking, and ranking. It is generally acknowledged that the model used in the pre-ranking stage must strike a balance between efficacy and efficiency. Thus, the most commonly employed architecture is the representation-focused vector product based model. However, this architecture lacks effective interaction between the query and document, resulting in a reduction in the effectiveness of the search system. To address this issue, we present a novel pre-ranking framework called \textbf{RankDFM}. Our framework leverages DeepFM as the backbone and employs a pairwise training paradigm to learn the ranking of videos under a query. The capability of RankDFM to cross features provides significant improvement in offline and online A/B testing performance. Furthermore, we introduce a learnable feature selection scheme to optimize the model and reduce the time required for online inference, equivalent to a tree model. Currently, RankDFM has been deployed in the search system of a shortvideo App, providing daily services to hundreds of millions users.
\end{abstract}




\maketitle

\section{Introduction}
\begin{figure}
  \centering
  \includegraphics[width=0.2\textwidth]{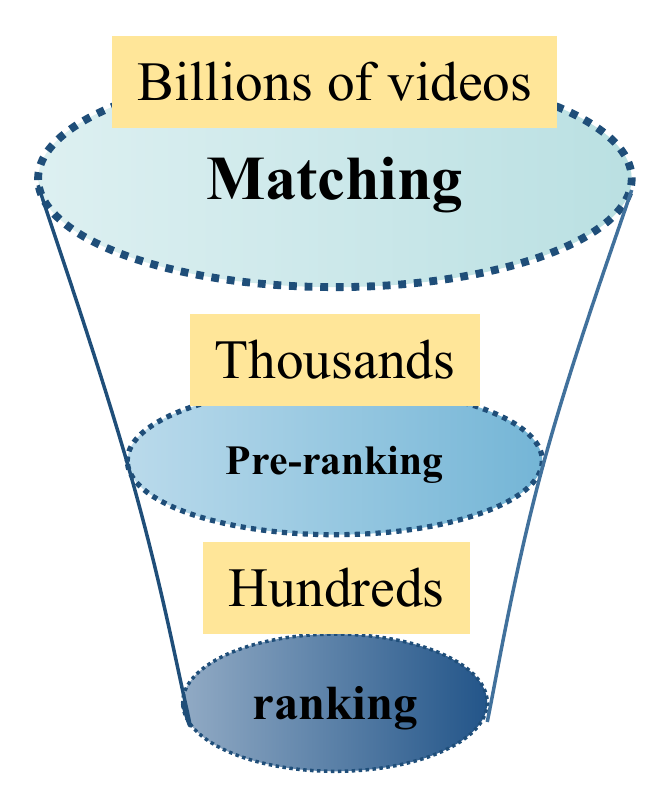}
  \caption{Illustration of cascade architecture for industrial information retrieval system.}
  \Description{Illustration of cascade architecture for industrial information retrieval system.}
  \label{cascade}
    \vspace{-14pt}
\end{figure}

As a result of the exponential proliferation of internet services in recent years, users have been grappling with an enormous volume of information. To mitigate this challenge, major internet platforms have employed search systems to aid billions of users in finding the information they require. The ranking system constitutes a pivotal aspect of these search systems, as it judiciously chooses a limited number of items from a massive pool of candidates. Regrettably, the intricate and elaborate ranking models are often hindered by the stringent constraint of low system latency, rendering them incapable of completing the ranking process for the entire candidate set.\cite{DBLP:conf/kdd/LiZXHCCX22, DBLP:conf/sigir/MaWZLZLLXZ21, DBLP:conf/recsys/CovingtonAS16, DBLP:journals/corr/abs-1905-06874, DBLP:conf/kdd/FanGZMSL19, DBLP:conf/kdd/PiBZZG19,DBLP:conf/aaai/LyuDHR20,DBLP:conf/kdd/ZouZCMCWSCY21}  To circumvent this limitation, the vast majority of ranking systems adopt a multi-stage cascade architecture, which is comprised of three linear sequential modules: matching, pre-ranking, and ranking. This architecture is depicted in Fig \ref{cascade}. 

The pre-ranking stage in the cascade architecture of ranking systems is tasked with filtering the items selected in the matching stage and forwarding them to the ranking module. This stage, unlike the ranking module that only needs to rank items on ten orders of magnitude items, often requires the processing of hundreds of items. To ensure system efficiency, simple models such as representation-based models \cite{DBLP:conf/cikm/HuangHGDAH13, DBLP:journals/corr/abs-1812-01190} are usually employed by the pre-ranking module, although these models reduce the overall effectiveness of the system due to limitations in their expressive power. On the other hand, interaction-based models, which allow for interactions among features, possess higher expressiveness compared to representation-based \cite{DBLP:journals/corr/abs-2007-16122} models. However, these models typically require high computational resources that are not affordable for search systems, resulting in the adoption of simpler models such as shallow MLP models \cite{DBLP:conf/kdd/FanGZMSL19,DBLP:journals/corr/abs-2007-16122}, which compromise the effectiveness of interaction-based pre-ranking modules. Hence, this paper sets out to study an interaction-based pre-ranking model that balances both efficiency and effectiveness.

In this paper, we present a new pre-ranking algorithm that balances efficiency and effectiveness for practical search systems. Our proposed approach, named RankDFM, incorporates a robust feature interaction information extraction model, specifically DeepFM \cite{10.5555/3172077.3172127}, as its backbone. The training phase employs the pairwise training strategy and generates positive and negative samples based on the ranking model's output. Consequently, the objective of RankDFM is to learn the item list ordering of the ranking model. During inference, RankDFM calculates the ranking score for an item directly from its features. Nevertheless, using DeepFM as the dominant model for the pre-ranking stage would be disastrous for the entire system due to its tremendous model complexity. To address this issue, we introduce the GDP (Gates with Differentiable Polarization) mechanism \cite{DBLP:conf/iccv/GuoYTWYL21} during training, which enables the automatic learning of necessary feature interactions for model pruning, effectively reducing the complexity of the model. Our experimental results, comprising both online A/B tests and offline evaluations, demonstrate the superiority of RankDFM over other baseline methods on two real-world datasets.

The contributions of our paper are as follows:
\begin{itemize}
\item We present a novel pre-ranking framework, RankDFM, which leverages DeepFM as its backbone to optimize feature interaction and enhance the ranking capability of the pre-training process.
\item To ensure that RankDFM's model complexity is commensurate with the demands of the pre-ranking stage, we incorporate the GDP (Gates with Differential Polarization) pruning mechanism, resulting in a significant reduction in computing resources while preserving RankDFM's performance.
\item Our extensive experiments on real-world datasets, coupled with an online A/B test , provide robust evidence of the superiority of our proposed RankDFM.
\end{itemize}
At present, the RankDFM framework has been successfully deployed in a shortvideo App search system, providing daily services to hundreds of millions of users.

\section{The Proposed Approach}
The pre-ranking stage is located in the middle module of the whole search cascade architecture. Its role is to rank thousands of items output by the matching layer, and send hundreds of items that rank first to the ranking layer for more accurate and sophisticated ranking. In other words, the function of the pre-ranking layer is to filter the results of the matching layer to relieve the pressure of the ranking layer. Formally, given a query $q$, suppose that the set of items returned by the matching layer is $\mathcal{S}_\mathcal{P}=\{\mathcal{I}_1,\mathcal{I}_2,\cdots,\mathcal{I}_n\}$, the role of the pre-ranking layer is to learn a function: $f=prerank (\mathcal{I}, q,  \mathcal{I}\times q)$ ($\mathcal{I}\in \mathcal{S}_\mathcal{P}$, $\mathcal{I}\times q$  represents the intersection features of item and query), score each item $\mathcal{I}$ in $\mathcal{S}_\mathcal{P}$, and select the top 100 items among them to deliver to the ranking layer.

\begin{figure*}
  \centering
  \includegraphics[width=0.8\textwidth]{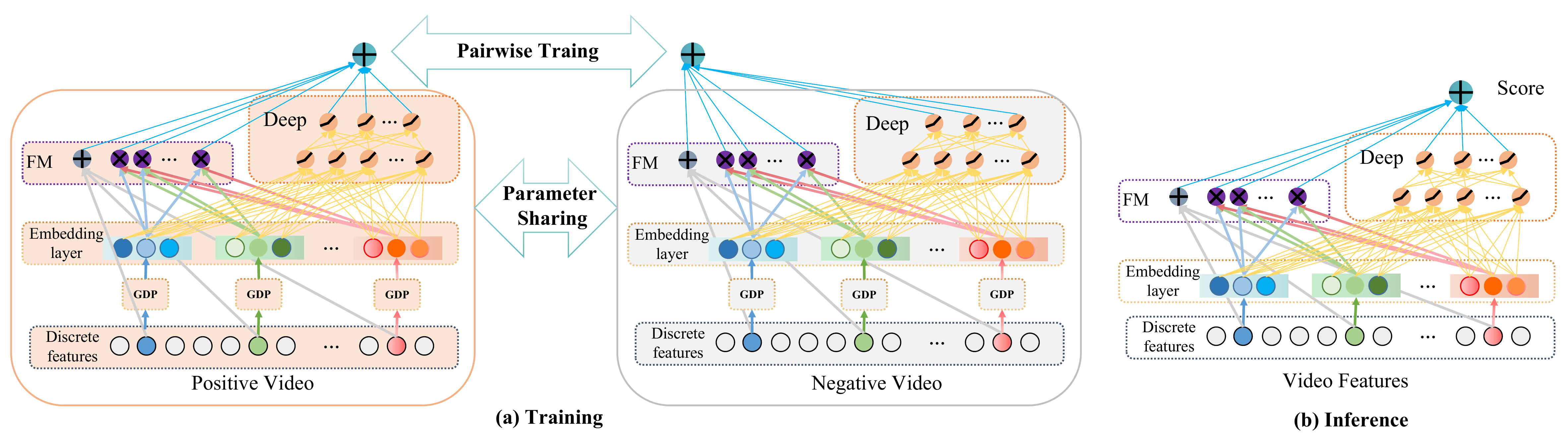}
  \caption{Overall training and inference framework for RankDFM.}
  \label{framework}
  \vspace{-10pt}
\end{figure*}

In this paper, we propose a novel pre-ranking framework RankDFM with both efficiency and effectiveness. The utilization of DeepFM as its backbone enhances the model's effectiveness, while the incorporation of the GDP pruning strategy optimizes its computational efficiency. Subsequently, a comprehensive overview of the training and inference details of RankDFM will be provided. A visual representation of the RankDFM framework is depicted in Fig.~\ref{framework}.

\subsection{Training of RankDFM}
As mentioned above, the learning goal of the pre-ranking stage is the ranking result of the candidates by the ranking module. Generally, there are three paradigms for learning to rank: Pointwise, pairwise, and listwise. To reduce training difficulty and improve training efficiency, we chose pairwise, a more economical training method. Next, we describe the specifics of training.

\subsubsection{Construct Training Samples}
The pairwise paradigm aims to construct partially ordered pairs of samples and learn to rank from comparisons. Thus, creating the sample pairs needed for training is the initial step. Using hundreds of candidates from the pre-ranking layer to learn the ranking results of the ranking layer is a straightforward method, yet this method suffers from SSB (Sample selection Bias), which makes the whole sample space inaccessible to the model. Therefore, our approach is to call the ranking model to score each item in the candidate set $\mathcal{S}_\mathcal{M}$ of the matching layer to obtain the set $\mathcal{S}_\mathcal{M}^R$. Then construct partial sample pairs on the set $\mathcal{S}_\mathcal{M}^R$.

Specifically, taking an item $\mathcal{I}$ in the set $\mathcal{S}_\mathcal{M}^R$ as an example, we tried the following three schemes:
\begin{itemize}
\item \textbf{\emph{Between-Levels.}} We divide all items into four levels according to the score interval of the ranking layer. Suppose that $\mathcal{I}$ belongs to the third level, then $\mathcal{I}$ select a certain number of items from the item sets of the first, second, and fourth levels, and does not select samples from the third level where it is located to construct partially ordered sample pairs.
\item \textbf{\emph{All-Levels.}} Similar to the previous one, all items are divided into four levels. Suppose that $\mathcal{I}$ belongs to the third level, $\mathcal{I}$ selects a certain number of items from the item set of the first, second, third, and fourth levels respectively to construct partially ordered sample pairs. This sample construction method takes into account the hard samples at the same level.
\item \textbf{\emph{Random.}} Although the previous method can well consider the hard samples, it still has one defect: this method needs to carefully regulate the proportion of anchor samples sampled at different levels, which is time-consuming and labor-intensive. In this context, we adopt the method of randomly selecting negative samples. Given an item $\mathcal{I}$ and a threshold value $\epsilon$, a certain number of samples are randomly selected from the item set whose difference in ranking layer scores is greater than the threshold $\epsilon$ to construct partially ordered sample pairs.
\end{itemize}
Finally, the third method discards the concept of level, and it is simple and easy to operate without fine-tuning the number of samples selected by items at different levels. We chose this method to construct partially ordered sample pairs. In conclusion, given a query $q$, we construct partially ordered sample pairs for its corresponding set $\mathcal{S}_\mathcal{M}^R$ according to the third approach.

\subsubsection{Feature Processing}
It is well knowledge that real-world search data streams contain a significant amount of missing data and outliers. We attempted numerous popular techniques to mitigate this problem, such 0-value filling, average value filling, deletion, etc., but they all proved unproductive owing to the complexity of the sample feature space. The fact that distinct features of a video may be subject to different distributions and that the data intervals are varied makes the task much more complicated. As a result, processing many aspects in a coherent manner is challenging. To alleviate these negative effects, fine processing of features is necessary.
Specifically, First of all, to solve the problem of inconsistent feature distribution, we adopt different normalizations for features of various distributions, such as maximum and minimum normalization for uniform distribution features, logarithm-based normalization for long-tail distribution features, etc., to standardize all feature values to the 0-1 interval, which is convenient for subsequent further processing. Next, we implemented a feature discretization mechanism to reduce the disturbance of outliers and missing values during model training. Moreover, this mechanism could also enhance the generalization ability of the model. In short, we put outliers and missing values into different buckets respectively, and then discretize the normal eigenvalues with equal width $w$, where width $w$ is an adjustable hyper-parameter.

\subsubsection{Model Training}
For the purpose of reducing parameters and adapting the online inference process, in the training phase, we adopted DeepFM as the backbone to build a parameter sharing siamese network. Suppose that there are many pairs of partially ordered sample pairs $\mathcal{P}=<\mathcal{I} _p, \mathcal{I} _n>$, during the training process, the positive DeepFM tower will score the positive sample $\mathcal{I} _p$ in the partially ordered sample pair $\mathcal{P}$, while the negative DeepFM tower will score the negative sample $\mathcal{I} _n$ in the partially ordered sample pair $\mathcal{P}$:
\begin{align}
\label{score}
  & S_p = Pos\_DeepFM(\mathcal{I} _p, q, \mathcal{I} _p \times q)\\
  & S_n = Neg\_DeepFM(\mathcal{I} _n, q, \mathcal{I} _n \times q)
\end{align}
where the parameters of the positive DeepFM tower $Pos\_DeepFM$ and the negative DeepFM tower $Neg\_DeepFM$ are shared.

For specific partial order sample pairs, our training goal is to make the positive DeepFM tower score positive samples higher than the negative DeepFM tower score negative samples. Furthermore, the model could learn the partial order relationship of all videos in a query $q$. Formally:
\begin{equation}
   \label{pairwise_loss}
    \mathcal{L}_{pair} = min(max(S_n-S_p+\sigma),0)
\end{equation}
where $\mathcal{L}_{pair}$ is the training objective based on the pairwise training paradigm. Simultaneously, in order to alleviate the excessive fluctuation of ranking scores of model and fix the scoring within a meaningful range, we also introduced a pointwise loss to let RankDFM learn the scores of the ranking layer. Finally, our training objective is as follows:
\begin{align}
\label{loss}
  & \mathcal{L} = \alpha \mathcal{L}_{pair} + (1-\alpha) \mathcal{L}_{point}\\
  & \mathcal{L}_{point} = min((S_n-r_n)^2+(S_p-r_p)^2)
\end{align}
where $r_n$ and $r_p$ are the scores of the ranking layer for negative sample and positive sample, respectively.

\subsubsection{Model Pruning}
\begin{figure}[h]
  \centering
  \includegraphics[width=0.35\textwidth]{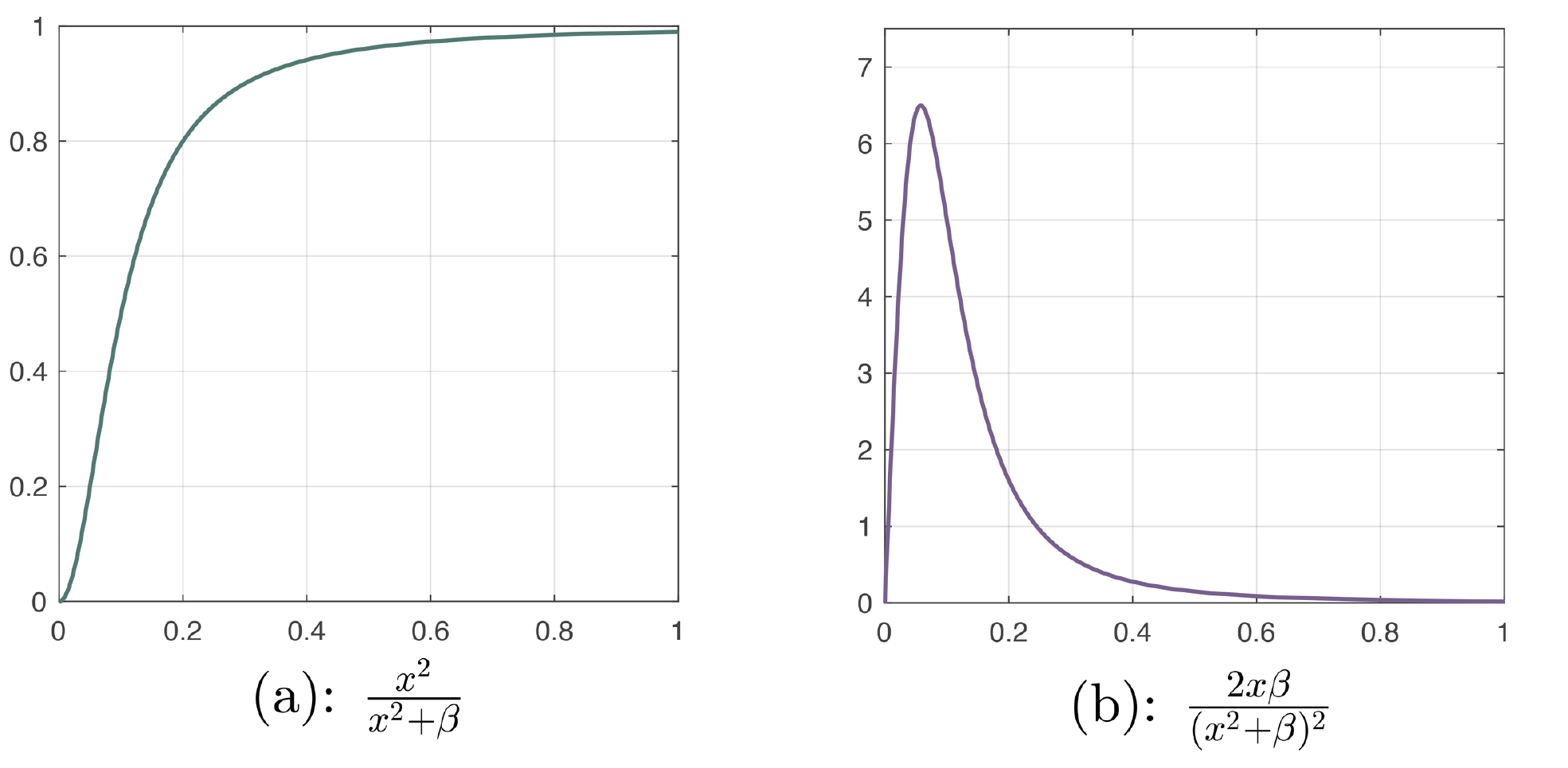}
  \caption{The graph of $g_{\beta}(x)$ and $g^{'}_{\beta}(x)$. $\beta=0.01$}
  \label{functions}
  \vspace{-14pt}
\end{figure}

The pre-ranking stage requires not only effectiveness but also efficiency. We use DeepFM as the backbone to build a pre-ranking model to ensure the effectiveness of the model. However, the high complexity of DeepFM also makes the pre-ranking stage overwhelmed, so model pruning is extremely necessary. We adopted the GDP mechanism for pruning. Specifically, GDP uses $g_{\beta}(x)$ as the polarization function because of the good characteristics of $g_{\beta}(x)$: the function value is an exact 0,1 value and could only stay stable at the 0,1 point according to the derivative $g^{'}_{\beta}(x)$ of $g_{\beta}(x)$ w.r.t. $x$ (As shown in Fig. \ref{functions}). We plug $g_{\beta}(x)$ as a gating function into the second-order and higher-order interaction modules of the model, and after one training epoch, we can obtain feature interactions with $g_{\beta}(x)$ value of 0, that is, unimportant interactions.
\begin{equation}
   \label{g}
    g_{\beta}(x) = \frac{x^2}{x^2+\beta}, g^{'}_{\beta}(x) = \frac{2x\beta}{(x^2+\beta)^2}
\end{equation}
where $\beta$ is a sufficiently small number. After we remove unimportant interactions from the model, the efficiency of the model is significantly improved, and the effectiveness of the model is not significantly reduced. 

\subsection{Inference of RanKDFM}
In the online inference stage, we do not need to construct partial order sample pairs, but only need to feed the sample features into the model for scoring, to facilitate the ranking and filtering in the pre-ranking stage.

\section{Experiments}
\subsection{Dataset}
We introduced two datasets to evaluate the performance of RankDFM: Kse0820 and Kse0720. To assess RankDFM's overall performance on the long tail and random queries, each dataset is split into two sub-datasets, random and long-tail. Each sub-dataset contains 5000 distinct inquiries and the recall outcomes of the matching layer that correspond to each query. Each dataset contains roughly 10 million videos.

\subsection{Experiment Settings}
\subsubsection{Baselines}
\begin{itemize}
    \item \textbf{\emph{RankNet.}} A conventional probability-based pairwise ranking algorithm.
    \item \textbf{\emph{Gbrank.}} A pairwise ranking algorithm based on XGBoost.
    \item \textbf{\emph{COLD.}} A DNN-based pre-ranking framework proposed by the Alibaba Group.
\end{itemize}

\subsubsection{Metric}
\textbf{\emph{Recall@N.}} The coincidence ratio of the top 600 items sorted by the pre-ranking model to the top $N$ items given by the ranking model. $N=10, 50, 600.$
\subsubsection{Parameter Configuration}
Here we give some detailed parameter configurations. The structure of the Deep side of the DeepFM is $input\times128\times64\times32\times16\times1$, the embedding size of features is $3$, the width $w$ of the bucket is $0.02$, the value of parameter $\alpha$ used to balance different losses is $0.9$, and the learning rate is $0.0001$. 

\subsection{Experiment Results}
As shown in Tab.~\ref{results}, our proposed RankDFM achieves state-of-the-art performance on both datasets. In addition, we also conduct comparative experiments on several variants of RankDFM, and the results are shown in Tab.~\ref{nopoint_result}. It is noticeable that the Pointwise loss we proposed can constrain the model scoring and w the model's effectiveness, while model's efficacy after being pruned by the GDP mechanism has not been materially diminished.

\subsection{Sensitivity Analysis}
\begin{figure}
  \centering
  \includegraphics[width=0.35\textwidth]{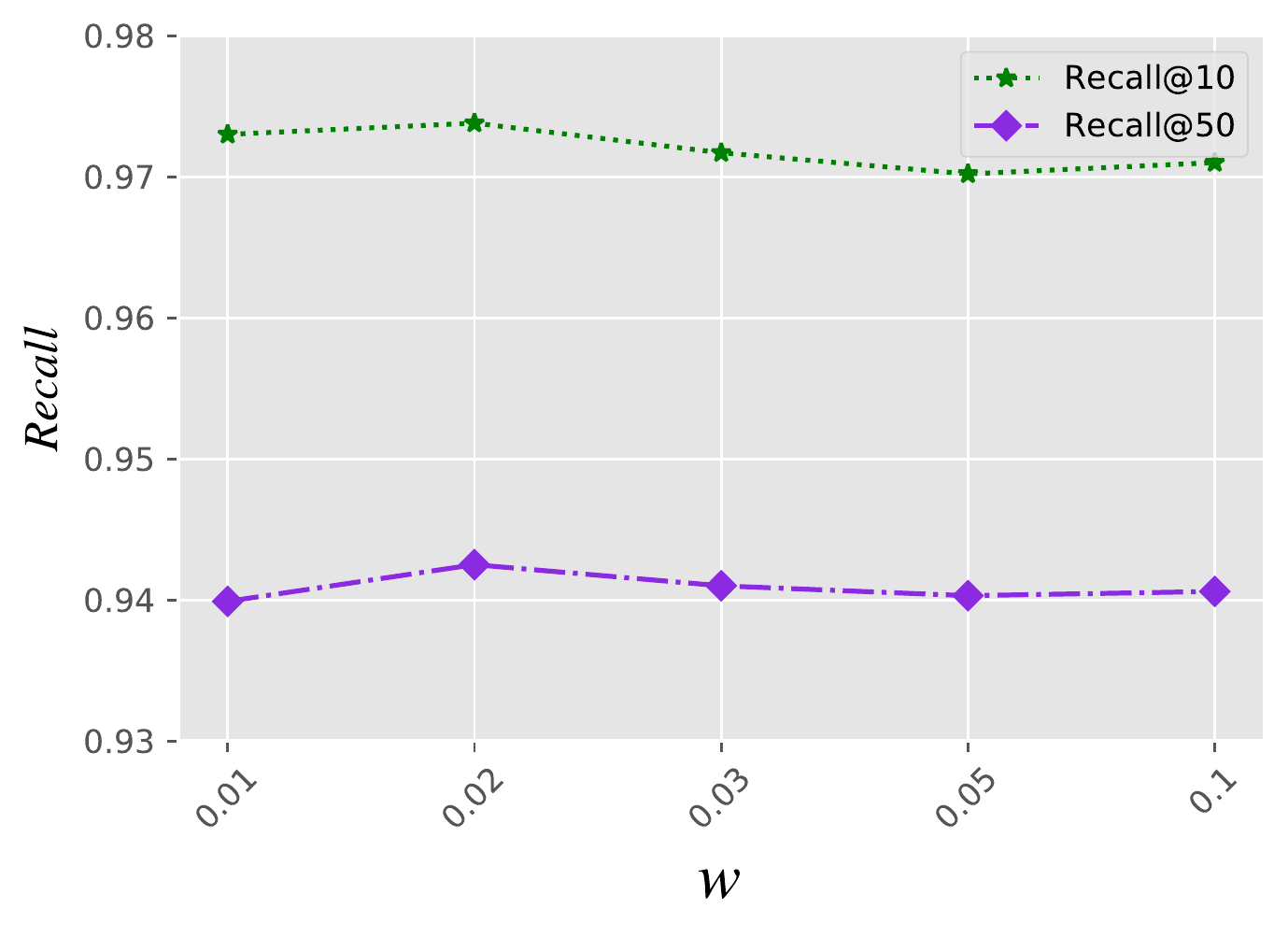}
  \caption{Illustration of cascade architecture for industrial information retrieval system.}
  \Description{Illustration of cascade architecture for industrial information retrieval system.}
  \label{sensi}
\end{figure}

We conduct sensitivity experiments on the parameter $w$ on the Kse0820 dataset. As shown in the Fig.~\ref{sensi}, the experimental results show that the model is stable to the parameter $w$, and the model achieves the best performance when $w=0.02$.

\subsection{Online A/B Test}
We conducted an online A/B test for 7 days to verify the effectiveness and efficiency of RankDFM. From the two key indicators of playback times and HPCR (Home Page Click Rate), we could see that our model has brought significant positive benefits to the entire system, and the two efficiency indicators of CPU usage and delay are the same as the base model.

\begin{table}[]
\caption{Performance comparison of different methods on Pre-ranking.}
\begin{tabular}{c|c|cc|cc}
\hline
\multirow{2}{*}{Baseline} & \multirow{2}{*}{Metric} & \multicolumn{2}{c|}{Kse0820}                               & \multicolumn{2}{c}{Kse0720}                               \\ \cline{3-6}
                          &                         & \multicolumn{1}{l}{random} & \multicolumn{1}{l|}{longtail} & \multicolumn{1}{l}{random} & \multicolumn{1}{l}{longtail} \\ \hline
\multirow{3}{*}{Ranknet}  & Recall@10               & 0.9217                     & 0.9116                       & 0.9122                     & 0.8991                       \\
                          & Recall@50               & 0.8976                     & 0.8528                       & 0.8790                     & 0.8488                       \\
                          & Recall@600              & 0.6246                     & 0.5981                       & 0.5998                     & 0.5726                       \\ \hline
\multirow{3}{*}{Gbrank}   & Recall@10               & 0.9566                     & 0.9316                       & 0.9457                     & 0.9229                       \\
                          & Recall@50               & 0.9217                     & 0.8847                       & 0.9068                     & 0.8765                       \\
                          & Recall@600              & 0.6521                     & 0.6202                       & 0.6231                     & 0.6078                       \\ \hline
\multirow{3}{*}{COLD}     & Recall@10               & 0.9662                     & 0.9442                       & 0.9615                     & 0.9396                       \\
                          & Recall@50               & 0.9381                     & 0.8925                       & 0.9227                     & 0.8916                       \\
                          & Recall@600              & 0.6684                     & 0.6351                       & 0.6420                     & 0.6213                       \\ \hline
\multirow{3}{*}{RankDFM}  & Recall@10               & \textbf{0.9738}            & \textbf{0.9507}              & \textbf{0.9679}            & \textbf{0.9443}              \\
                          & Recall@50               & \textbf{0.9444}            & \textbf{0.9080}              & \textbf{0.9292}            & \textbf{0.8991}              \\
                          & Recall@600              & \textbf{0.6753}            & \textbf{0.6412}              & \textbf{0.6493}            & \textbf{0.6288}              \\ \hline
\end{tabular}
\label{results}
\end{table}

\begin{table}[]
\caption{Performance comparison of different Variants of RankDFM.}
\begin{tabular}{c|c|cc|cc}
\hline
\multirow{2}{*}{Method}                                                       & \multirow{2}{*}{Metric} & \multicolumn{2}{c|}{Kse0820}                                         & \multicolumn{2}{c}{Kse0720}                                         \\ \cline{3-6} 
                                                                              &                         & \multicolumn{1}{l|}{random}          & \multicolumn{1}{l|}{longtail} & \multicolumn{1}{l|}{random}          & \multicolumn{1}{l}{longtail} \\ \hline
\multirow{3}{*}{\begin{tabular}[c]{@{}c@{}}RankDFM\\ (no point)\end{tabular}} & Recall@10               & \multicolumn{1}{c|}{0.9721}          & 0.9495                        & \multicolumn{1}{c|}{\textbf{0.9679}} & 0.9431                       \\
                                                                              & Recall@50               & \multicolumn{1}{c|}{\textbf{0.9444}} & 0.9058                        & \multicolumn{1}{c|}{0.9270}          & 0.8969                       \\
                                                                              & Recall@600              & \multicolumn{1}{c|}{0.6737}          & 0.6388                        & \multicolumn{1}{c|}{0.6468}          & \textbf{0.6288}              \\ \hline
\multirow{3}{*}{\begin{tabular}[c]{@{}c@{}}RankDFM\\ (GDP)\end{tabular}}      & Recall@10               & \multicolumn{1}{c|}{0.9729}          & 0.9501                        & \multicolumn{1}{c|}{0.9660}          & 0.9439                       \\
                                                                              & Recall@50               & \multicolumn{1}{c|}{0.9420}          & 0.9069                        & \multicolumn{1}{c|}{0.9278}          & 0.8984                       \\
                                                                              & Recall@600              & \multicolumn{1}{c|}{0.6745}          & 0.6403                        & \multicolumn{1}{c|}{0.6481}          & 0.6255                       \\ \hline
\multirow{3}{*}{RankDFM}                                                      & Recall@10               & \multicolumn{1}{c|}{\textbf{0.9738}} & \textbf{0.9507}               & \multicolumn{1}{c|}{0.9664}          & \textbf{0.9443}              \\
                                                                              & Recall@50               & \multicolumn{1}{c|}{0.9425}          & \textbf{0.9080}               & \multicolumn{1}{c|}{\textbf{0.9292}} & \textbf{0.8991}              \\
                                                                              & Recall@600              & \multicolumn{1}{c|}{\textbf{0.6753}} & \textbf{0.6412}               & \multicolumn{1}{c|}{\textbf{0.6493}} & 0.6261                       \\ \hline
\end{tabular}
\label{nopoint_result}
\end{table}

\begin{table}[]
\caption{ Online effectiveness and efficiency of the proposed pre-ranking model and base pre-ranking model.}
\begin{tabular}{cccccc}
\cline{1-5}
Model                                                   & HPCR     & Views    & CPU(\%) & latency(ms) &  \\ \cline{1-5}
base                                                    & +0\%        & +0\%       &   60      & 3            &  \\
RankDFM                                                 & /         & /         &  80       & 15            &  \\
\begin{tabular}[c]{@{}c@{}}RankDFM (GDP)\end{tabular} & +0.127\% & +0.398\% &  65       &  5           &  \\ \cline{1-5}
\end{tabular}
\label{online}
\end{table}

\section{Conclusion}
In this paper, we propose RankDFM, a unique efficient and effective pre-ranking framework. RankDFM aims to utilize the DeepFM model as the backbone, which can enhance the model's feature cross ability, increase its efficacy, and build siamese DeepFM towers for partial rank learning based on the pairwise training paradigm. Meanwhile, in order to overcome the low efficiency of DeepFM model, we adopted the GDP pruning mechanism to prune the model, reducing the inference time of the model. The effectiveness of the RankDFM proposed by us is proved by sufficient experiments both offline and online.


\bibliographystyle{ACM-Reference-Format}
\bibliography{rankdfm}


\begin{thebibliography}{13}


\ifx \showCODEN    \undefined \def \showCODEN     #1{\unskip}     \fi
\ifx \showDOI      \undefined \def \showDOI       #1{#1}\fi
\ifx \showISBNx    \undefined \def \showISBNx     #1{\unskip}     \fi
\ifx \showISBNxiii \undefined \def \showISBNxiii  #1{\unskip}     \fi
\ifx \showISSN     \undefined \def \showISSN      #1{\unskip}     \fi
\ifx \showLCCN     \undefined \def \showLCCN      #1{\unskip}     \fi
\ifx \shownote     \undefined \def \shownote      #1{#1}          \fi
\ifx \showarticletitle \undefined \def \showarticletitle #1{#1}   \fi
\ifx \showURL      \undefined \def \showURL       {\relax}        \fi
\providecommand\bibfield[2]{#2}
\providecommand\bibinfo[2]{#2}
\providecommand\natexlab[1]{#1}
\providecommand\showeprint[2][]{arXiv:#2}

\bibitem[Chen et~al\mbox{.}(2019)]%
        {DBLP:journals/corr/abs-1905-06874}
\bibfield{author}{\bibinfo{person}{Qiwei Chen}, \bibinfo{person}{Huan Zhao},
  \bibinfo{person}{Wei Li}, \bibinfo{person}{Pipei Huang}, {and}
  \bibinfo{person}{Wenwu Ou}.} \bibinfo{year}{2019}\natexlab{}.
\newblock \showarticletitle{Behavior Sequence Transformer for E-commerce
  Recommendation in Alibaba}.
\newblock \bibinfo{journal}{\emph{CoRR}}  \bibinfo{volume}{abs/1905.06874}
  (\bibinfo{year}{2019}).
\newblock
\showeprint[arXiv]{1905.06874}
\urldef\tempurl%
\url{http://arxiv.org/abs/1905.06874}
\showURL{%
\tempurl}


\bibitem[Covington et~al\mbox{.}(2016)]%
        {DBLP:conf/recsys/CovingtonAS16}
\bibfield{author}{\bibinfo{person}{Paul Covington}, \bibinfo{person}{Jay
  Adams}, {and} \bibinfo{person}{Emre Sargin}.}
  \bibinfo{year}{2016}\natexlab{}.
\newblock \showarticletitle{Deep Neural Networks for YouTube Recommendations}.
  In \bibinfo{booktitle}{\emph{Proceedings of the 10th {ACM} Conference on
  Recommender Systems, Boston, MA, USA, September 15-19, 2016}}.
  \bibinfo{publisher}{{ACM}}, \bibinfo{pages}{191--198}.
\newblock
\urldef\tempurl%
\url{https://doi.org/10.1145/2959100.2959190}
\showDOI{\tempurl}


\bibitem[Fan et~al\mbox{.}(2019)]%
        {DBLP:conf/kdd/FanGZMSL19}
\bibfield{author}{\bibinfo{person}{Miao Fan}, \bibinfo{person}{Jiacheng Guo},
  \bibinfo{person}{Shuai Zhu}, \bibinfo{person}{Shuo Miao},
  \bibinfo{person}{Mingming Sun}, {and} \bibinfo{person}{Ping Li}.}
  \bibinfo{year}{2019}\natexlab{}.
\newblock \showarticletitle{{MOBIUS:} Towards the Next Generation of Query-Ad
  Matching in Baidu's Sponsored Search}. In
  \bibinfo{booktitle}{\emph{Proceedings of the 25th {ACM} {SIGKDD}
  International Conference on Knowledge Discovery {\&} Data Mining, {KDD} 2019,
  Anchorage, AK, USA, August 4-8, 2019}}. \bibinfo{publisher}{{ACM}},
  \bibinfo{pages}{2509--2517}.
\newblock
\urldef\tempurl%
\url{https://doi.org/10.1145/3292500.3330651}
\showDOI{\tempurl}


\bibitem[Guo et~al\mbox{.}(2017)]%
        {10.5555/3172077.3172127}
\bibfield{author}{\bibinfo{person}{Huifeng Guo}, \bibinfo{person}{Ruiming
  Tang}, \bibinfo{person}{Yunming Ye}, \bibinfo{person}{Zhenguo Li}, {and}
  \bibinfo{person}{Xiuqiang He}.} \bibinfo{year}{2017}\natexlab{}.
\newblock \showarticletitle{DeepFM: A Factorization-Machine Based Neural
  Network for CTR Prediction}. In \bibinfo{booktitle}{\emph{Proceedings of the
  26th International Joint Conference on Artificial Intelligence}}.
  \bibinfo{publisher}{AAAI Press}, \bibinfo{pages}{1725–1731}.
\newblock


\bibitem[Guo et~al\mbox{.}(2021)]%
        {DBLP:conf/iccv/GuoYTWYL21}
\bibfield{author}{\bibinfo{person}{Yi Guo}, \bibinfo{person}{Huan Yuan},
  \bibinfo{person}{Jianchao Tan}, \bibinfo{person}{Zhangyang Wang},
  \bibinfo{person}{Sen Yang}, {and} \bibinfo{person}{Ji Liu}.}
  \bibinfo{year}{2021}\natexlab{}.
\newblock \showarticletitle{{GDP:} Stabilized Neural Network Pruning via Gates
  with Differentiable Polarization}. In \bibinfo{booktitle}{\emph{2021
  {IEEE/CVF} International Conference on Computer Vision, {ICCV} 2021,
  Montreal, QC, Canada, October 10-17, 2021}}. \bibinfo{publisher}{{IEEE}},
  \bibinfo{pages}{5219--5230}.
\newblock
\urldef\tempurl%
\url{https://doi.org/10.1109/ICCV48922.2021.00519}
\showDOI{\tempurl}


\bibitem[Huang et~al\mbox{.}(2013)]%
        {DBLP:conf/cikm/HuangHGDAH13}
\bibfield{author}{\bibinfo{person}{Po{-}Sen Huang}, \bibinfo{person}{Xiaodong
  He}, \bibinfo{person}{Jianfeng Gao}, \bibinfo{person}{Li Deng},
  \bibinfo{person}{Alex Acero}, {and} \bibinfo{person}{Larry~P. Heck}.}
  \bibinfo{year}{2013}\natexlab{}.
\newblock \showarticletitle{Learning deep structured semantic models for web
  search using clickthrough data}. In \bibinfo{booktitle}{\emph{22nd {ACM}
  International Conference on Information and Knowledge Management, CIKM'13,
  San Francisco, CA, USA, October 27 - November 1, 2013}}.
  \bibinfo{publisher}{{ACM}}, \bibinfo{pages}{2333--2338}.
\newblock
\urldef\tempurl%
\url{https://doi.org/10.1145/2505515.2505665}
\showDOI{\tempurl}


\bibitem[Li et~al\mbox{.}(2022)]%
        {DBLP:conf/kdd/LiZXHCCX22}
\bibfield{author}{\bibinfo{person}{Xiang Li}, \bibinfo{person}{Xiaojiang Zhou},
  \bibinfo{person}{Yao Xiao}, \bibinfo{person}{Peihao Huang},
  \bibinfo{person}{Dayao Chen}, \bibinfo{person}{Sheng Chen}, {and}
  \bibinfo{person}{Yunsen Xian}.} \bibinfo{year}{2022}\natexlab{}.
\newblock \showarticletitle{AutoFAS: Automatic Feature and Architecture
  Selection for Pre-Ranking System}. In \bibinfo{booktitle}{\emph{{KDD} '22:
  The 28th {ACM} {SIGKDD} Conference on Knowledge Discovery and Data Mining,
  Washington, DC, USA, August 14 - 18, 2022}}. \bibinfo{publisher}{{ACM}},
  \bibinfo{pages}{3241--3249}.
\newblock
\urldef\tempurl%
\url{https://doi.org/10.1145/3534678.3539083}
\showDOI{\tempurl}


\bibitem[Lyu et~al\mbox{.}(2020)]%
        {DBLP:conf/aaai/LyuDHR20}
\bibfield{author}{\bibinfo{person}{Zequn Lyu}, \bibinfo{person}{Yu Dong},
  \bibinfo{person}{Chengfu Huo}, {and} \bibinfo{person}{Weijun Ren}.}
  \bibinfo{year}{2020}\natexlab{}.
\newblock \showarticletitle{Deep Match to Rank Model for Personalized
  Click-Through Rate Prediction}. In \bibinfo{booktitle}{\emph{The
  Thirty-Fourth {AAAI} Conference on Artificial Intelligence, {AAAI} 2020, The
  Thirty-Second Innovative Applications of Artificial Intelligence Conference,
  {IAAI} 2020, The Tenth {AAAI} Symposium on Educational Advances in Artificial
  Intelligence, {EAAI} 2020, New York, NY, USA, February 7-12, 2020}}.
  \bibinfo{publisher}{{AAAI} Press}, \bibinfo{pages}{156--163}.
\newblock
\urldef\tempurl%
\url{https://doi.org/index.php/AAAI/article/view/5346}
\showDOI{\tempurl}


\bibitem[Ma et~al\mbox{.}(2021)]%
        {DBLP:conf/sigir/MaWZLZLLXZ21}
\bibfield{author}{\bibinfo{person}{Xu Ma}, \bibinfo{person}{Pengjie Wang},
  \bibinfo{person}{Hui Zhao}, \bibinfo{person}{Shaoguo Liu},
  \bibinfo{person}{Chuhan Zhao}, \bibinfo{person}{Wei Lin},
  \bibinfo{person}{Kuang{-}Chih Lee}, \bibinfo{person}{Jian Xu}, {and}
  \bibinfo{person}{Bo Zheng}.} \bibinfo{year}{2021}\natexlab{}.
\newblock \showarticletitle{Towards a Better Tradeoff between Effectiveness and
  Efficiency in Pre-Ranking: {A} Learnable Feature Selection based Approach}.
  In \bibinfo{booktitle}{\emph{{SIGIR} '21: The 44th International {ACM}
  {SIGIR} Conference on Research and Development in Information Retrieval,
  Virtual Event, Canada, July 11-15, 2021}}. \bibinfo{publisher}{{ACM}},
  \bibinfo{pages}{2036--2040}.
\newblock
\urldef\tempurl%
\url{https://doi.org/10.1145/3404835.3462979}
\showDOI{\tempurl}


\bibitem[Pi et~al\mbox{.}(2019)]%
        {DBLP:conf/kdd/PiBZZG19}
\bibfield{author}{\bibinfo{person}{Qi Pi}, \bibinfo{person}{Weijie Bian},
  \bibinfo{person}{Guorui Zhou}, \bibinfo{person}{Xiaoqiang Zhu}, {and}
  \bibinfo{person}{Kun Gai}.} \bibinfo{year}{2019}\natexlab{}.
\newblock \showarticletitle{Practice on Long Sequential User Behavior Modeling
  for Click-Through Rate Prediction}. In \bibinfo{booktitle}{\emph{Proceedings
  of the 25th {ACM} {SIGKDD} International Conference on Knowledge Discovery
  {\&} Data Mining, {KDD} 2019, Anchorage, AK, USA, August 4-8, 2019}}.
  \bibinfo{publisher}{{ACM}}, \bibinfo{pages}{2671--2679}.
\newblock
\urldef\tempurl%
\url{https://doi.org/10.1145/3292500.3330666}
\showDOI{\tempurl}


\bibitem[Wang et~al\mbox{.}(2020)]%
        {DBLP:journals/corr/abs-2007-16122}
\bibfield{author}{\bibinfo{person}{Zhe Wang}, \bibinfo{person}{Liqin Zhao},
  \bibinfo{person}{Biye Jiang}, \bibinfo{person}{Guorui Zhou},
  \bibinfo{person}{Xiaoqiang Zhu}, {and} \bibinfo{person}{Kun Gai}.}
  \bibinfo{year}{2020}\natexlab{}.
\newblock \showarticletitle{{COLD:} Towards the Next Generation of Pre-Ranking
  System}.
\newblock \bibinfo{journal}{\emph{CoRR}}  \bibinfo{volume}{abs/2007.16122}
  (\bibinfo{year}{2020}).
\newblock
\showeprint[arXiv]{2007.16122}


\bibitem[Wu et~al\mbox{.}(2018)]%
        {DBLP:journals/corr/abs-1812-01190}
\bibfield{author}{\bibinfo{person}{Wenjin Wu}, \bibinfo{person}{Guojun Liu},
  \bibinfo{person}{Hui Ye}, \bibinfo{person}{Chenshuang Zhang},
  \bibinfo{person}{Tianshu Wu}, \bibinfo{person}{Daorui Xiao},
  \bibinfo{person}{Wei Lin}, {and} \bibinfo{person}{Xiaoyu Zhu}.}
  \bibinfo{year}{2018}\natexlab{}.
\newblock \showarticletitle{{EENMF:} An End-to-End Neural Matching Framework
  for E-Commerce Sponsored Search}.
\newblock \bibinfo{journal}{\emph{CoRR}}  \bibinfo{volume}{abs/1812.01190}
  (\bibinfo{year}{2018}).
\newblock
\showeprint[arXiv]{1812.01190}
\urldef\tempurl%
\url{http://arxiv.org/abs/1812.01190}
\showURL{%
\tempurl}


\bibitem[Zou et~al\mbox{.}(2021)]%
        {DBLP:conf/kdd/ZouZCMCWSCY21}
\bibfield{author}{\bibinfo{person}{Lixin Zou}, \bibinfo{person}{Shengqiang
  Zhang}, \bibinfo{person}{Hengyi Cai}, \bibinfo{person}{Dehong Ma},
  \bibinfo{person}{Suqi Cheng}, \bibinfo{person}{Shuaiqiang Wang},
  \bibinfo{person}{Daiting Shi}, \bibinfo{person}{Zhicong Cheng}, {and}
  \bibinfo{person}{Dawei Yin}.} \bibinfo{year}{2021}\natexlab{}.
\newblock \showarticletitle{Pre-trained Language Model based Ranking in Baidu
  Search}. In \bibinfo{booktitle}{\emph{{KDD} '21: The 27th {ACM} {SIGKDD}
  Conference on Knowledge Discovery and Data Mining, Virtual Event, Singapore,
  August 14-18, 2021}}. \bibinfo{publisher}{{ACM}},
  \bibinfo{pages}{4014--4022}.
\newblock
\urldef\tempurl%
\url{https://doi.org/10.1145/3447548.3467147}
\showDOI{\tempurl}


\end{thebibliography}

\end{document}